\documentclass[%
 aip,
rsi,%
 amsmath,amssymb,
preprint,%
]{revtex4-1}

\usepackage{graphicx}% Include figure files
\usepackage{dcolumn}% Align table columns on decimal point
\usepackage{bm}% bold math
\usepackage{float}
\usepackage{mathtools}
\DeclarePairedDelimiter{\abs}{\lvert}{\rvert}

\begin{document}

%\preprint{AIP/123-QED}

\title[Balanced XAS]{Balanced Detection in Femtosecond X-ray Absorption Spectroscopy to Reach the Ultimate Sensitivity Limit}% Force line breaks with \\

\newcommand{\LCLS}{Linear Coherent Light Source, SLAC National Accelerator Lab, 2575 Sand Hill Rd., Menlo Park, CA 94025, USA}

\newcommand{\SLAC}{SLAC National Accelerator Lab, 2575 Sand Hill Rd., Menlo Park, CA 94025, USA}

\newcommand{\UofA}{Van der Waals-Zeeman Institute, Institute of Physics, University of Amsterdam, Science Park 904, 1098 XH, Amsterdam, The Netherlands}

\newcommand{\MESA}{Faculty of Science and Technology and MESA+ Institute for Nanotechnology, University of Twente, P.O. Box 217, 7500 AE, Enschede, The Netherlands}

\newcommand{\SIMES}{Stanford Institute for Materials and Energy Science, SLAC National Accelerator Laboratory, 2575 Sand Hill Road, Menlo Park, California 94025, USA}

\newcommand{\FLASH}{FS-FLASH, Deutsches Elektronen-Synchrotron (DESY), Notkestr. 85 D-22607 Hamburg, Germany}

\newcommand{\XFEL}{Spectroscopy \& Coherent Scattering, European X-Ray Free-Electron Laser Facility GmbH, Holzkoppel 4, 22869 Schenefeld Germany}

\author{W.F. Schlotter}
 \email{wschlott@slac.stanford.edu}
 \affiliation{\LCLS}

\author{M. Beye}%
\affiliation{\FLASH}

\author{S. Zohar}%
\affiliation{\LCLS}
\author{G. Coslovich }%
\affiliation{\LCLS}
\author{G. L. Dakovski }%
\affiliation{\LCLS}
\author{M.-F. Lin}%
\affiliation{\LCLS}

\author{Y. Liu}%
\affiliation{\SLAC}

\author{A. Reid}%
\affiliation{\LCLS}

\author{S. Stubbs }%
\affiliation{\LCLS}
\altaffiliation{SWISS FEL, Paul Scherrer Institut, Forschungsstrasse 111, 5232 Villigen, Switzerland}

\author{P. Walter }%
\affiliation{\LCLS}

\author{K. Nakahara}%
\affiliation{\LCLS}

\author{P. Hart}%
\affiliation{\LCLS}

\author{P. S. Miedema}%
\affiliation{\FLASH}

\author{L. Le Guyader}%
\affiliation{\SLAC}
\affiliation{\XFEL}

\author{K. Hofhuis}%
\affiliation{\MESA}

\author{Phu Tran Phong Le}
\affiliation{\MESA}

\author{Johan E. ten Elshof}
\affiliation{\MESA}

\author{H. Hilgenkamp}%
\affiliation{\MESA}

\author{G. Koster}%
\affiliation{\MESA}

\author{X.H. Verbeek}%
\affiliation{\UofA}
%\altaffiliation{Materials Theory, ETH Zurich, Wolfgang-Pauli-Strasse 27, 8093 Zürich, Switzerland}

\author{S. Smit}%     
\affiliation{\UofA}

\author{M.S. Golden}%
\affiliation{\UofA}

\author{H.A. D\"{u}rr}%
\affiliation{Uppsala Universitet, Dept of Physics and Astronomy, Box 516, 751 20 Uppsala, Sweden}

\author{A. Sakdinawat }%
\affiliation{\SLAC}

\date{\today}% It is always \today, today,

\begin{abstract}
X-ray absorption spectroscopy (XAS) is a powerful and well established technique with sensitivity to elemental and chemical composition.\cite{Stoehr2003} Despite these advantages, its implementation has not kept pace with the development of ultrafast pulsed x-ray sources where XAS can capture femtosecond chemical processes.  X-ray Free Electron Lasers (XFELs) deliver femtosecond, narrow bandwidth ($\frac{\Delta E}{E} < 0.5\%$) pulses containing $\sim 10^{10}$ photons.\cite{Bostedt2016}  However, the energy contained in each pulse fluctuates thus complicating pulse by pulse efforts to quantify the number of photons.  Improvements in counting the photons in each pulse have defined the state of the art for XAS sensitivity.  Here we demonstrate a final step in these improvements through a balanced detection method that approaches the photon counting shot noise limit.   In doing so, we obtain high quality absorption spectra from the insulator-metal transition in VO$_2$ and unlock a method to explore dilute systems, subtle processes and nonlinear phenomena with ultrafast x-rays.  The method is especially beneficial for x-ray light sources where integration and averaging are not viable options to improve sensitivity.  
\end{abstract}

\maketitle

X-ray absorption spectroscopy (XAS) is a valuable tool for understanding electronic structures; it has impacted scientific insights in fields ranging from heterogeneous catalysis to correlated electron systems like those exhibiting metal-insulator transitions.\cite{Schloegl2015,Imada1998}  Soft X-ray absorption spectroscopy is sensitive to both elemental composition and chemical state.  It can be used to probe bonds formed by specific elements such as carbon, nitrogen and oxygen as well as to understand the role that d-orbital electrons play in the many emergent properties of transition metal compounds.\cite{Stoehr2003}  The high sensitivity that enabled these advances is provided by the stable and sizable x-ray photon flux (number of photons per second) generated by modern storage ring synchrotron light sources.

The advent of ultrafast pulsed x-ray sources sparked excitement in the promise of tracking electron dynamics with femtosecond resolution.\cite{Schoenlein2000}  Early demonstrations at low flux sources further fueled the enthusiasm.\cite{Cavalleri2005,Boeglin2010}  However, the development of XAS at next generation x-ray free electron laser (XFEL) sources has not been limited by the number of available photons but instead by methods for normalization.  The temporal structure of the x-ray pulse FEL sources, which deliver millijoule-scale pulses with tens of femtosecond durations, has hindered the development of XAS.  The deluge of photons from each pulse challenges the linearity and dynamic range of established detectors such as photodiodes and multi-channel plates.  Consequently the application of XAS at XFELs to time resolved measurements has been restricted, often requiring extensive experimental beamtime for limited, albeit significant, results. \cite{Kubin2017,Higley2016,Kroll2016} 

The scarcity of short pulses at FEL sources, which typically operate between 10-120 Hz, has favored spectroscopic methods that take advantage of the full pulse bandwidth.\cite{Bernstein2009} Improved normalization can be achieved using a transmission grating to generate two spectrally dispersed copies of the incident beam. \cite{Katayama2013}  Recently, off axis zone plate illumination was used to fill an area detector (e.g. CCD) in order to resolve ultrafast pump probe signals in a single shot.\cite{Buzzi2017} Because the spectrum for each pulse is different, improvements to signal quality require improved detection or averaging and careful normalization over multiple pulses.  However, the complete measurement of both energy spectrum and temporal evolution in a single pulse is not conducive to high sensitivity spectroscopy because of limitations in averaging due to detector readout noise.    
%% Need to add Gunther and Martin's OPTICS LETTERS to the approve paragraph 

By restricting each single shot measurement to a narrow spectral energy bandwidth recorded with superior normalization afforded by high dynamic range detection, one can obtain high fidelity absorption measurements that improve with averaging.  This is the standard method for XAS measurement at synchrotron storage rings where advances in grating monochromators enabled high spectral resolution XAS using detection methods including total electron yield, fluorescence and transmission.\cite{Stoehr2003,Strocov2010} This approach has been used to record the highest quality XAS spectra  at x-ray FEL sources.\cite{Kroll2016} Recently the use of a CCD to measure transmission greatly enhanced the sensitivity for time-resolved XAS measurements at a FEL. By expanding the x-ray beam on the area detector the low noise properties of the CCD are exploited while improving dynamic range. \cite{Higley2016}

Here we employ the ability of a grating to generate consistent copies of a beam combined with a zone plate to uniformly illuminate a high sensitivity CCD area detector, thus demonstrating photon counting noise as the main limitation to XAS sensitivity. 

Photon counting noise or shot noise represents the ultimate limit to photon detection sensitivity as described by statistical optics. Because each soft x-ray photon observed on a silicon detector generates hundreds of electrons, cooled, low noise electronics can easily resolve a single photon.  Therefore, the dynamic range in this regime is not limited by readout electronics, but rather the total number of detected photons which scales with the illuminated detector area. A typical megapixel CCD area detector can linearly observe 10$^{9}$ photons ($\lambda$=1 nm). \cite{Strueder2010} The uncertainty in counting because of photon shot noise is $\sqrt{N}$ or SNR=$10^4$ for $10^8$ detected photons.   However, a single shot sensitivity of $10^4$ is far from the current single shot state-of-the-art for XAS at an FEL.   

Sensitivity at (or near) the photon counting limit will extend XAS to study dilute samples and open the possibility to measure subtle changes associated with non-linear processes.  It represents the most efficient data collection method, therefore it is also ideally suited for radiation sensitive samples.

\begin{figure*}[t]
  
    \includegraphics[width=7in]{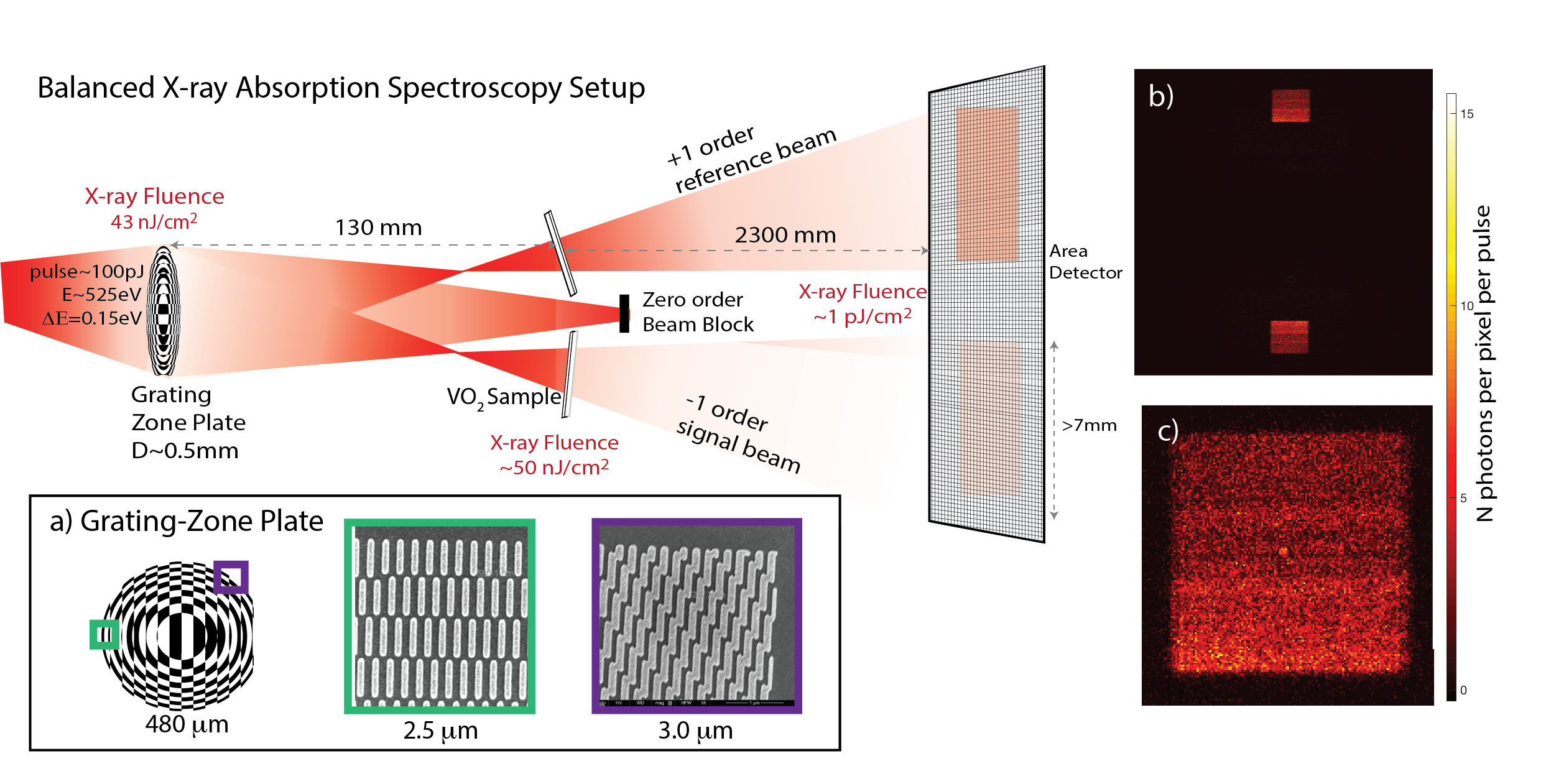}
    \caption{The two diffracted beams form the reference, $r$, that does not pass through the sample and the signal, $s$, that does.  a) The grating zone plate structure is fabricated by electron beam lithography (see methods) and contains a 600 nm outer zone width.   Because the average single pulse fluence illuminating the zone plate is 43 nJ/cm$^2$ the zone plate is well below the damage threshold. b) A pnCCD destructively measures both beams from a single pulse as they are detected within isolated regions of interest (ROI). c) The reference ROI containing 5.4$\times$10$^4$ photons spread over 16,510 pixels is shown with a graininess characteristic of photon counting noise.      }
    
    \label{fig:setup}
\end{figure*}

Our experimental arrangement is illustrated in Figure \ref{fig:setup} where monochromatic, soft x-rays ($\lambda$=2.4 nm, $\frac{\lambda}{\Delta\lambda}$=3500) illuminate a transmission diffraction grating.  The transmission diffractive optical element includes a zone plate and grating combined on a single structure.  In this way two identical, balanced, highly divergent beams are generated by amplitude division.  Even if the two beams differ because of imperfection in the diffractive element, their photon number ratio will be consistent.  Upon detection, the number of photons in each of the first order beams differs based on a Poisson statistical parent distribution. For the XAS measurement one beam passes through the sample (signal, \emph{s}) while the other serves as a reference (\emph{r}) beam. Because of the divergence of the zone plate focal length ($f$=122 mm for a photon energy of 525 eV), the beams expand rapidly to fill the large area detector.  Following photon detection of the two beams on the CCD, the detector is read out and corrected (see methods) and the two regions of interest (ROI) are selected for integration, see Figure \ref{fig:setup}.  The transmission through the sample is calculated simply by computing the ratio of the two integrated regions of interest.
  
To demonstrate shot noise limited performance we validate the setup without a sample in place to form two identical beams.  

\begin{figure}[t]
  
    \includegraphics[width=3.5in]{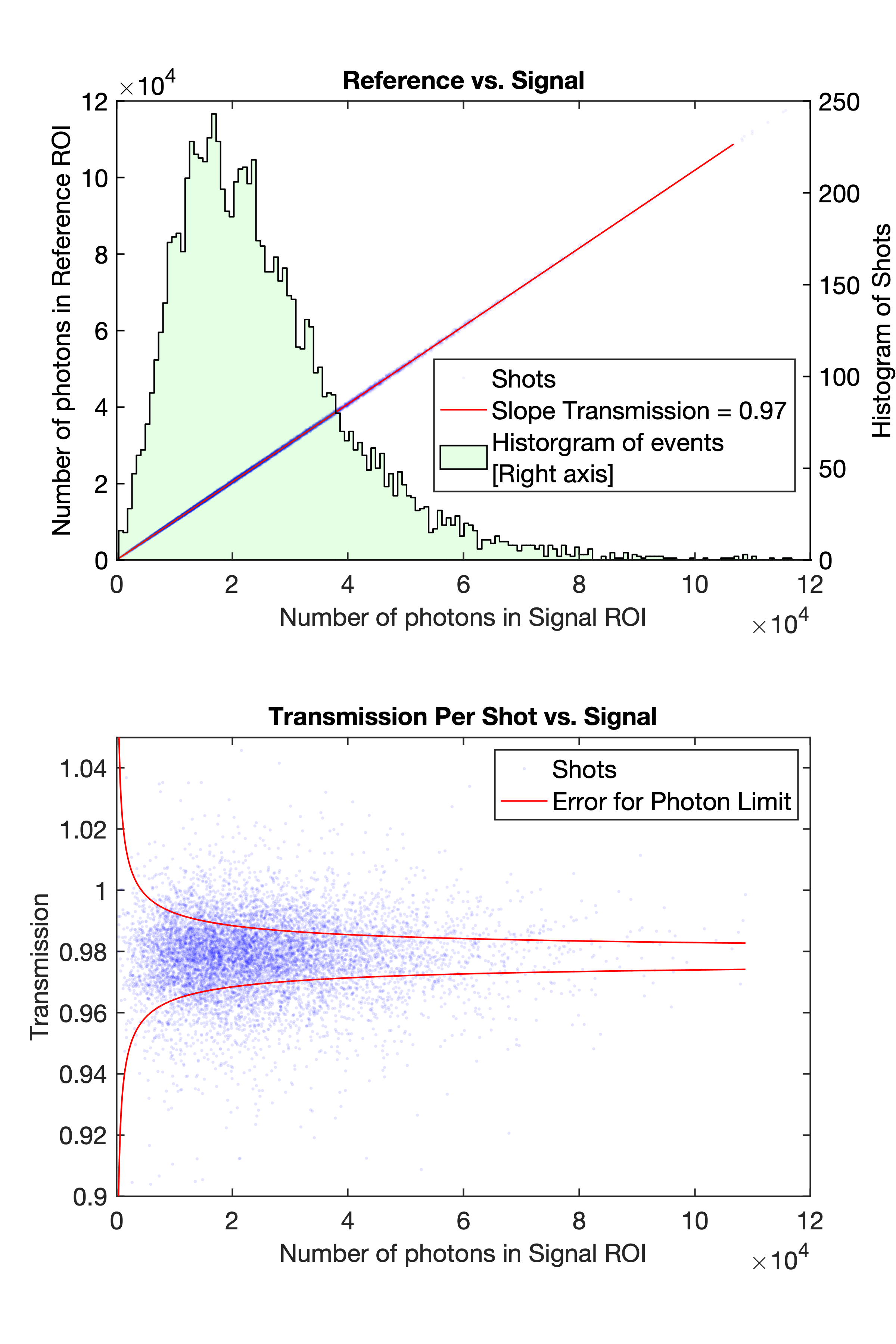}
    \caption{Data for the control case where no sample is inserted into the the signal beam.  a) Correlation between signal and reference (integrated ROIs) for 8,739 single events (blue dots).  The red line is a fit to the data where the slope represents the transmission. b) Transmission vs. signal for each shot are plotted.  Solid lines indicate one standard error of the mean or the photon limited error:  $T_{lim}=T_{mean}\pm \sqrt{2/N}$ where N is the number of photons measured in the signal ROI.  In the purely photon noise limited case, $2\sigma$ (68.2\%) of the shots would fall between the error curves.  Over the full signal range shown 62.4\% of the recorded shots are within this photon limited envelope.}
    
    \label{fig:sigVref}
\end{figure}

Figure \ref{fig:sigVref} a) shows a correlation plot between the two ROIs.    A line fit to the correlation gives the ratio of the number of photons in the two beams which is 0.97$\pm$0.001.  The fit describes the transmission which should ideally be unity.   Therefore, the fit, in this case, is the systematic error to the ratio between the two beams. By analysing the ratio of the two ROIs vs. the reference ROI it becomes clear that the fidelity of the measurement increases with the number of photons in each pulse as illustrated in Figure \ref{fig:sigVref}b). The fraction of recorded events within the photon limited error envelope increases as the number of detected photons in the signal increases.  

To quantify the sensitivity of this method we calculate the Signal to Noise Ratio (SNR) as the ratio of the mean transmission, $T$, to the standard deviation of the transmission $\sigma_T$ such that $SNR=T/\sigma_{T}$.

\begin{figure}[t]
    \includegraphics[width=3.5in]{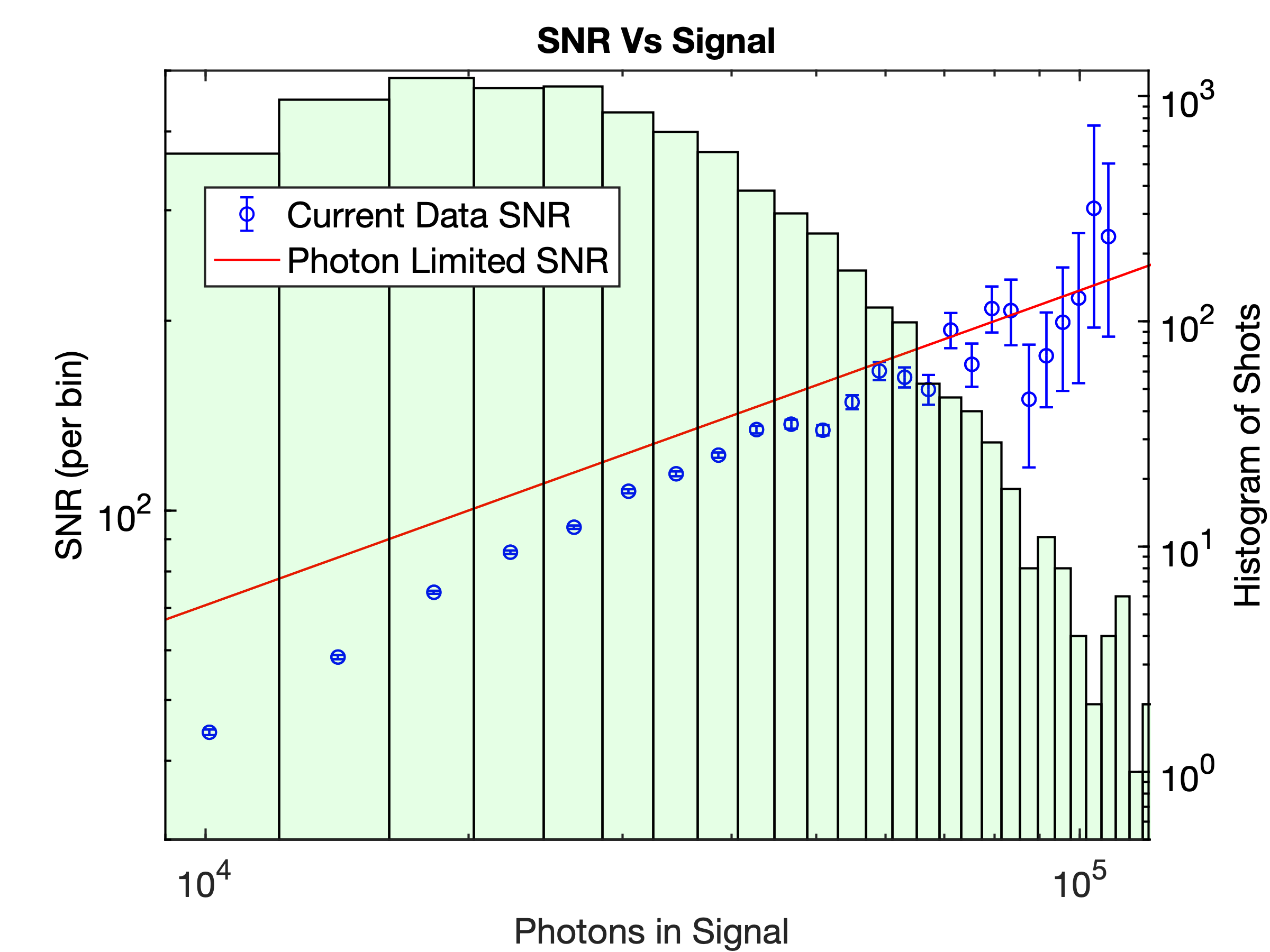}
    \caption{The SNR (left axis) data points with errorbars are plotted vs. the signal level in photons.  The errorbars represent the standard error which is $\frac{\sigma_{SNR}}{\sqrt{N}}$ where $N$ is the number of measurement shots per point.  The photon limited SNR (solid line) is provided to illustrate the maximum possible SNR.  For shots collecting more than 6$\times$10$^4$ photons the SNR reaches the photon noise limit.  The histogram (right axis) of shaded bars displays the number of shots used to calculate the SNR and error.  }
    \label{fig:SNR_plot}
\end{figure}

The SNR improves as the number of detected photons increases and Figure \ref{fig:SNR_plot} shows the intensity dependence.   For comparison the SNR in the photon counting noise limit, $ SNR = \sqrt{\frac{N}{2}}$, is plotted as well.   The error bars on the SNR ($\sigma_{SNR}$) are derived by propagating error as follows:

\begin{equation}
V(\sigma_T^2) = \frac{(n-1)^2}{n^3}\bigg(\mu_4-\frac{n-3}{n-1}\sigma_T^4\bigg)
\label{from_rao_var_of_var}
\end{equation}
where $V(\sigma_T^2)$ is the variance of the standard deviation of the transmission, $\mu_q$ is the $q^{th}$ order moment of $T$ and $n$ is the number of data points per bin.\cite{Rao2001}  Using error propagation we can arrive at the standard deviation of the SNR, $\sigma_{SNR}$. 

\begin{equation}
\sigma_{SNR} = \frac{\sqrt{V(\sigma_T^2)}}{2\sigma_T} \abs*{\frac{<T>}{\sigma_T^2}}
\label{eq_SNR}
\end{equation}
%The errorbars in Fig. \ref{fig:SNR_plot} represent the standard error which is $\frac{\sigma_{SNR}}{\sqrt{N}}$ where $N$ is the number of measurements.   

We see that the SNR falls short of the shot noise limit by 35\% at low signal and approaches it for the highest signal shots. The explanation for the closing of this gap may come from detector readout noise, like fixed pattern noise and errors in the common mode correction.

\begin{figure}[b]
    \centering
    \includegraphics[width=3.5in]{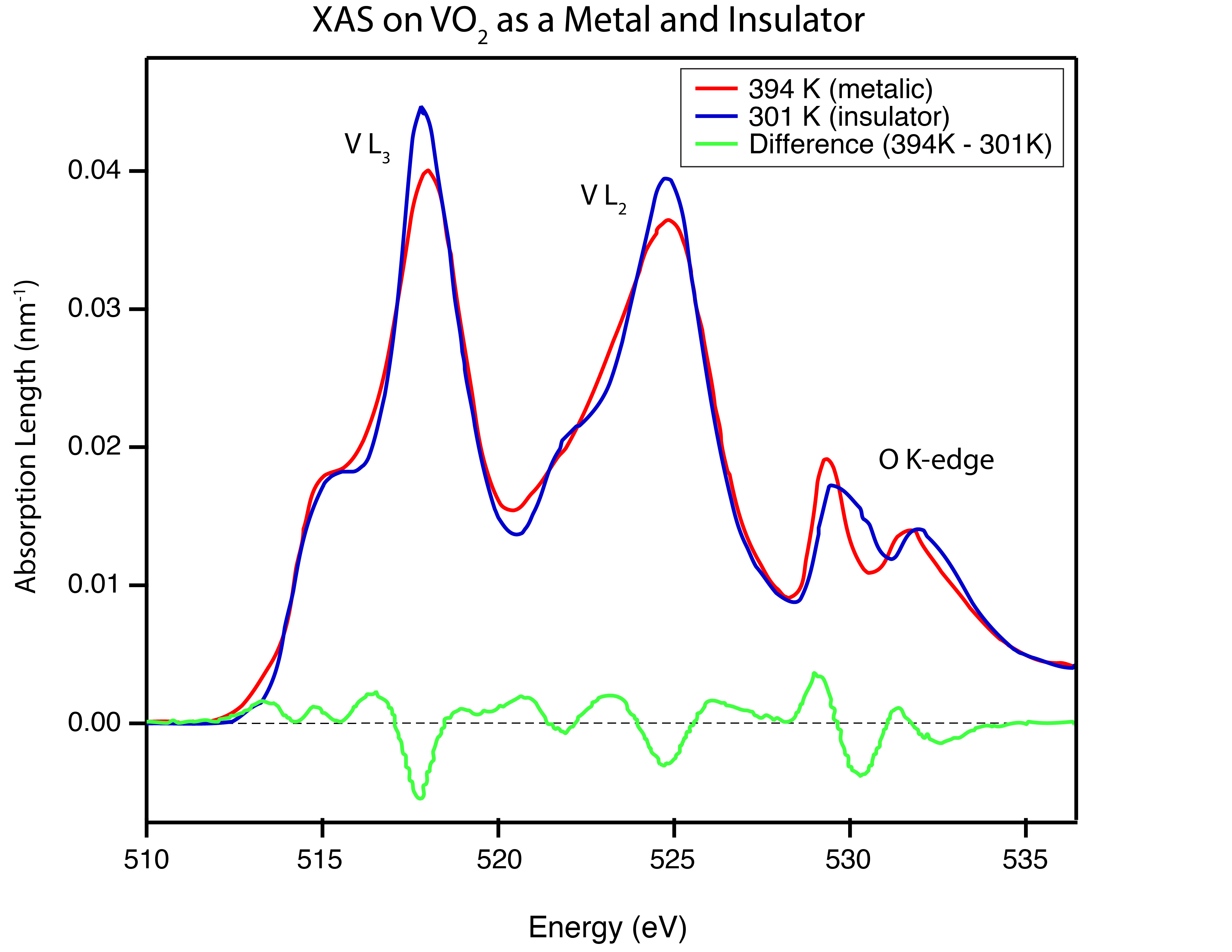}
    \caption{Absorption spectra recorded on the 50 nm thick VO$_2$ sample at temperatures above (394 K) and below (301 K) the insulator-to-metal transition.  The spectra span the vanadium L-edges and the oxygen K-edge.  The difference between the two spectra is plotted and depicts the changes in the unoccupied electronic states.}
    \label{fig:VO2}
\end{figure}

To demonstrate the advantage of this spectroscopy method we recorded an XAS spectrum of the often-studied transition metal oxide, VO$_2$. Vanadium dioxide is technologically interesting because of its insulator-to-metal transition slightly above room temperature.  The vanadium L-edges and oxygen K-edge are within a 50 eV photon energy scan range.   While the ROI in the balanced beam detection method shift spatially during the 50 eV photon energy scan, they do remain fully on the detector.  To generate these spectra, the sample transmission is calculated by taking the ratio of the sum of the detected photons in the sample ROI with respect to the reference ROI.  The absorption length, $\alpha$, corresponds to: $$ \alpha =- \frac{ln{(T)}}{t_{sample}}$$ where $T$ is the sample transmission and $t_{sample}$ is the thickness.  The method affords a simple, yet absolute, calculation of the absorption length because the sample is studied in transmission and an identical reference is recorded.  Figure \ref{fig:VO2} shows two XAS spectra for VO$_2$ measured at temperatures above and below the metal-to-insulator transition.   Each spectrum was recorded in less than 17 minutes at 120 Hz repetition rate for a total of 82 nJ of incident x-ray energy on the sample over each entire scan.  These are the highest quality VO$_2$ spectra recorded using ultrafast x-ray pulses and are comparable to spectra recorded at synchrotron storage rings for similar samples and conditions. \cite{Aetukuri2013,Gray2016,Le2019} The difference between the two spectra that is plotted in Figure \ref{fig:VO2} is also validated by storage ring lightsource spectra.

VO$_2$ is particularly sensitive to changes in temperature so it is important to minimize the x-ray fluence incident on the sample.  In this case the spot size was 80  $\mu$m at 500 eV and 30 $\mu$m at 550 eV resulting in an average per pulse fluence of 10 nJ/cm$^2$ and 75 nJ/cm$^2$ respectively. The pulse duration was 110 fs.  Because the distance between the zone plate and the sample remained fixed during the scan we see this change in fluence as a result of the longitudinal dependence, of the zone plate focal length on wavelength. The spot size on the sample can be increased simply by moving it toward the detector because of the large angular divergence.  However, because the total integrated energy for the measured spectra was only 82 nJ, x-ray induced changes are, in fact, undetectable. 

Moving the sample position closer to the focus enables higher intensity and because the focal spot can be less than 25 nm it is conceivable to reach a fluence of $\sim 1 $kJ/cm$^2$ at an FEL source.  Such intensities are unprecedented at a narrow bandwidth, thus opening new possibilities to study x-ray induced non-linear absorption phenomena. \cite{Stoehr2019}   

%Discussion 

The method presented scales with photon counting noise.  Therefore an optimized setup capable of illuminating a full area detector (2.5 cm$^2$) could detect 10$^9$ soft x-ray photons (fewer at shorter wavelengths) and thus realize a per shot SNR of 20,000.  Such sensitivity would be suitable for measuring subtle differences in adsorption from dilute solutions or very thin samples or interfaces. Because the position of the sample can be used to control the fluence and the method is in the photon noise regime this represents the most efficient XAS measurement possible.   This is important at high intensity pulsed sources as well as low intensity pulsed sources where the measurement of every photon is essential for obtaining the best possible data quality.   The experimental geometry affords sufficient space to introduce a magnetic field or a pulsed laser to optically pump the sample.  

We have demonstrated x-ray absorption spectroscopy limited only by photon counting statistics.  This method is well suited for pulsed sources such as x-ray free electron lasers like the one used here.   Moreover, thanks to this efficiency, samples prone to change or damage upon x-ray illumination can be explored at the lowest possible exposure fluence.  Our exemplary XAS spectra of the metal-insulator transition in VO$_2$ validate that the balanced beam detection is accurate, fast and robust.  Application of balanced beam XAS will provide new opportunities for time resolved x-ray experiments.

\section{Methods}

\subsection{X-ray Parameters}
Data were collected on the Soft X-ray Instrument for Materials Science (SXR) at the Linac Coherent Light Source (LCLS).   For these spectroscopic energy scans the wavelength was continuously adjusted by scanning the SXR monochromator between 510 eV and 550 eV. \cite{Schlotter2012,Heimann2011}  The LCLS electron beam energy was concurrently scanned to maintain the maximum transmission through the monochromator.   To ensure the zone plate was illuminated with a clean wavefront, the exit slits on the monochromator were closed to 5 $\mu$m and the KB mirror system was set to focus in the horizontal 1 m downstream of the zone plate.   Consequently the zone plate was illuminated and overfilled by a 1.3 x 1.9 mm (h x v) spot. 

\subsection{Zone plate parameters and fabrication }
The integrated beam splitting zone plate was fabricated with gold on a 100 nm thick SiN membrane using electron beam lithography and electroplating. See the micrograph in Figure \ref{fig:setup} a).  The diameter of the zone plate was 480 $\mu$m with an outer zone width of 600 nm while the grating period was 225 nm. The combination of the zone plate and grating into a single optic predicts a diffraction efficiency of 4.5$\%$ into the first order beams. \cite{Chang2006}  The parameters for the integrated zone plate grating were optimized to maximize the spot size on the detector as well as the distance between the two beams as they intersect the sample plane.

\subsection{Sample and Reference Geometry }
Because 90\% of the illuminating beam is transmitted by the zoneplate, care was taken to fully attenuate it using a beamstop located 180 mm downstream of the zone plate optic.   The horizontal focus of the KB mirror is downstream of the beamstop, thus preventing the direct beam from illuminating and saturating the detector. 

For the VO$_2$ measurements the sample and reference were positioned 130 mm downstream of the zone plate at which point the spacing between the two diffraction orders was 3 mm.  The reference was a commercially available 3 mm diameter 200 $\mu$m thick Si substrate with a 200 nm thick Si$_3$N$_4$ membrane to form a 250 x 250 $\mu$m window at the center.    The sample was a 50 nm thick VO$_2$ film grown on an identical Si$_3$N$_4$ system.  \cite{Le2019}  The Si substrate between the two Si$_3$N$_4$ windows serves as an order sorting aperture for the zone plate focal orders. 

\subsection{Detection and Photon Calibration}
To optically isolate the detector, a 200 nm thick Al film was introduced 650 mm downstream of the zoneplate.  The detector plane was located 2428 mm downstream of the zoneplate.  The pnCCD consists of two halves, each consisting of 1024 x 512 pixels each with a size of 75 $\mu$m. The two detector halves were separated by 1.4 mm.   The pnCCD was operated in high gain mode (1.1 ADU/eV) where the noise level is 0.12 ADU when the detector temperature is -55 C, thus providing single photon sensitivity.   The illumination levels were below the full well depth of 300 000 electrons, ensuring no distortion from saturation.  \cite{Strueder2010}   

The sensitivity of this measurement requires careful attention to background subtraction, gain correction and common mode correction. \cite{Blaj2015}  The background was subtracted after averaging 3480 images collected from the detector under the same conditions at which the data was recorded.   To compensate for pixel by pixel variations in gain, a correction matrix is applied to each pixel as determined via flat field illumination.  Finally, a common mode correction was applied to compensate for the time-dependent variation in amplifier gain.  For this it was important to mask the area of illumination by the signal and reference ROI.  

Determining the ADU per incident photon from the detector is crucial for correctly evaluating the photon counting noise limit correctly. A histogram of the counts per pixel was generated for various regions of interest using 1500 collected images. From these histograms the first photon peak coincides with 574$\pm$61 ADU per photon.

\begin{acknowledgments}
Use of the Linac Coherent Light Source (LCLS), SLAC National Accelerator Laboratory, is supported by the U.S. Department of Energy, Office of Science, Office of Basic Energy Sciences under Contract No. DE-AC02-76SF00515. L.L.G. would like to thank the VolkswagenStiftung for the financial support through the Peter-Paul-Ewald Fellowship. P.M. and M.B. acknowledge funding the Helmholtz Association via grant VH-NG-1105. X.H.V., S.S, M.G., K.H., H.H. and G.K. acknowledge the NWO/FOM programme DESCO (VP149), which is financed by the Netherlands Organisation for Scientific Research (NWO).  P.T.P.L., J.E.tE.  and G.K. acknowledge the NWO/CW ECHO grant ECHO.15.CM2.043. P.T.P.L acknowledges financial support from the Netherlands Organization for Scientific Research (NWO) in the framework of the Chemical Sciences ECHO programme.  The authors kindly thank Daniel Higley for reviewing the manuscript.  

\end{acknowledgments}

\section*{Author Contributions}

W.F.S., A.S. and H.D. conceived the experiment. M.B., G.C., G.L.D., H.D., K.H., L.L.G., M.F.L., Y.L., P.M., A.R., A.S., W.F.S., S.S., P.W., S.Z., S.S., K.N., P.A.H and X.H.V. planned and participated in the experiment.   Y.L, A.S and W.F.S. designed and fabricated the grating zone plate structures.  M.G., H.H., G.K., K.H., W.F.S., S.S., P.T.P.L. and J.E.tE. designed, fabricated and characterized the  VO$_2$ sample systems.  L.L.G, W.F.S, S.Z. and X.H.V. performed data analysis and model development.  The manuscript preparation was lead by W.F.S with the participation of all co-authors.  

\section*{Competing financial interests}
The authors declare no competing financial interests.

%\nocite{*}
\bibliography{x229_ref_040320}% Produces the bibliography via BibTeX.

\end{document}